\documentclass[aps,showpacs,pra,superscriptaddress,]{revtex4}
\usepackage{amsmath}
\usepackage{amsfonts}
\usepackage{amssymb}
\usepackage{bm}
\usepackage{graphicx}

\begin{document}

\title{Wave-speed management of dipole, bright and W-shaped solitons in
optical metamaterials}
\author{ Houria Triki}
\affiliation{Radiation Physics Laboratory, Department of Physics, Faculty of Sciences,
Badji Mokhtar University, P. O. Box 12, 23000 Annaba, Algeria}
\author{Vladimir I. Kruglov}
\affiliation{Centre for Engineering Quantum Systems, School of Mathematics and Physics,
The University of Queensland, Brisbane, Queensland 4072, Australia}

\begin{abstract}
Wave-speed management of soliton pulses in a nonlinear metamaterial
exhibiting a rich variety of physical effects that are important in a wide
range of practical applications, is studied both theoretically and
numerically. Ultrashort electromagnetic pulse transmission in such
inhomogeneous system is described by a generalized nonlinear Schr\"{o}dinger
equation with space-modulated higher-order dispersive and nonlinear effects
of different nature. We present the discovery of three types of periodic
wave solutions that are composed by the product of Jacobi elliptic functions
in the presence of all physical processes. Envelope solitons of the dipole,
bright and W-shaped types are also identified, thus illustrating the
potentially rich set of localized pulses in the system. We develop an
effective similarity transformation method to investigate the soliton
dynamics in the presence of the inhomogeneities of media. The application of
developed method to control the wave speed of the presented solitons is
discussed. The results show that the wave speed of dipole, bright and
W-shaped solitons can be effectively controlled through spatial modulation
of the metamaterial parameters. In particular, the soliton pulses can be
decelerated and accelerated by suitable variations of the distributed
dispersion parameters.
\end{abstract}

\maketitle
\affiliation{}

\section{Introduction}

Soliton formation in nonlinear metamaterials is presently a very active area
of research \cite{PLi,Pendry,Reed,Zharov,Parazzoli,Shen}. Such new materials
are known as negative-index materials and more commonly referred to as
left-handed metamaterials \cite{Kozyrev1}. To demonstrate solitons
experimentally, left-handed nonlinear transmission lines, employed as
nonlinear metamaterials, have been recently used to study the generation of
envelope solitons \cite{Kozyrev1,Kozyrev2,Kozyrev3,Kozyrev4}. In addition,
the stable generation of soliton pulses has been also demonstrated
experimentally in an active nonlinear metamaterial formed by a left-handed
transmission line inserted into a ring resonator \cite{Kozyrev3}. Moreover,
dark envelope solitons in a practical left-handed nonlinear transmission
line with series nonlinear capacitance are demonstrated by circuit analysis,
which showed that the left-handed nonlinear transmission lines could support
dark solitons by tailoring the circuit parameters \cite{Kozyrev4}.
Furthermore, a left-handed nonlinear electrical lattice has been shown to
support the formation of discrete envelope solitons of the bright and dark
type \cite{English}. These significant results indicate that the soliton
propagation is one of the physically relevant phenomena associated with
practical nonlinear metamaterials.

To describe the transmission of an ultrashort optical pulse through an
homogeneous nonlinear metamaterial, a generalized nonlinear Schr\"{o}dinger
equation (NLSE) for a dispersive dielectric susceptibility and permeability
has been introduced by Scalora et al. \cite{Scalora}. One should note here
that the utilization of the NLSE to describe the nonlinear wave dynamics is
not only restricted to nonlinear metamaterials, but also to other important
physical media like optical fibers \cite{Agra}, Bose-Einstein condensates 
\cite{Beitia}, plasma physics \cite{Dodd}, biomolecular dynamics \cite%
{Davydov}, etc. Regarding realistic optical waveguiding media, they are
actually inhomogeneous because of the existence of some nonuniformities
which arise due to various factors such as the variation of lattice
parameters in the optical medium, the imperfection of manufacture and
fluctuation of the system diameters \cite{Abdullaev,Lei}. With consideration
of the inhomogeneities in an optical material, the theoretical description
of the light pulse dynamics is mainly based on the generalized NLSE equation
with varying group velocity dispersion, nonlinearity, and gain (absorption)
coefficients \cite{Serkin,Kruglov3,Kruglov4}. Especially in optics, the
application of such generalized NLSE has stimulated further studies of the
integrable inhomogeneous equations giving rise to the concepts of
nonautonomous and self-similar solitons \cite%
{Serkin,Kruglov3,Kruglov4,Chen,Ponomarenko,Serkin2}. For the soliton
transmission in an optical medium within the femtosecond duration range,
however, the higher-order effects influenced by the variations of material
parameters should be also taken into account \cite{Abdullaev1}.

From a physical viewpoint, a soliton structure can be described by four
parameters which are the frequency (or velocity), amplitude (or width), time
position and phase \cite{Has5}. An appropriate modulation of these
parameters allows efficient control of the soliton dynamics. Recently, the
problem of optimal control the parameters of optical solitons which is also
called the problem of soliton management becomes a subject of significant
interest \cite{Belya,R1,R2}. This because concepts of soliton dispersion
management and soliton control in optical fiber systems constitute a
physically relevant developments in the practical application of envelope
solitons for optical transmissions \cite%
{Hasegawa1,Hasegawa2,Hasegawa3,Hasegawa4}. Interestingly, studies of
dispersion management have demonstrated that several effects can be reduced
by utilizing this technique such as the modulational instability \cite%
{Doran1}, Gordon-Haus effect resulting from the interaction with noise \cite%
{Doran2}, radiation due to lumped amplifiers compensating the fiber loss 
\cite{Doran3}, and time jitters caused by the collisions between signals 
\cite{Doran4}.

While the control of solitons shape or amplitude under dispersion and
nonlinearity managements have been demonstrated in physical systems within
the framework of both cubic and higher-order NLSE models, the control of
soliton wave speed in nonlinear media has been studied thus far only for the
cubic NLSE case in \cite{Luke}. With use of this envelope equation which
includes only two physical effects, it is shown that one can control the
wave speed of bright and dark NLS solitons by appropriate modification of
the dispersion and nonlinearity coefficients \cite{Luke}. A challenging
problem is the study of wave-speed management of solitons in the presence of
higher-order effects, which come into play as pulse durations get shorter
and peak powers increase. Also, a more significant issue is to examine the
wave-speed management of various types of solitons under the contributions
of these higher-order processes. As is known, in addition to the bright and
dark soliton types, solitons could also display other complex shapes such as
dipole and W-shaped structures. It is worth noting that dipole-mode solitons
which are consisting of two peaks \cite{Susanto}, have been recently
observed in a three-level cascade atomic system where it has been
experimentally demonstrated that the key to observe them is to create via
Kerr nonlinearity, a high enough index contrast in the atomic medium by
laser-induced index gratings \cite{Yanpeng}. As concerns solitons that take
the shape of W, these have been firstly presented in optical fibers
described by the higher-order NLSE with third-order dispersion,
self-steepening, and self-frequency shift effects in \cite{Zhoou}. Up to the
present time, the control of soliton wave speed in optical metamaterials
supporting higher-order effects has not been reported to our knowledge. In
this paper, we present the analysis of wave-speed management of dipole,
bright and W-shaped solitons in an inhomogeneous optical metamaterial
exhibiting not only the group velocity dispersion and self-phase-modulation,
but also a rich variety higher-order effects. One should note that studying
on the control of soliton wave speed is not only of scientific relevance but
also of practical significance.

This paper is organized as follows. In Sec. II, we present the generalized
higher-order NLSE that governs the few-cycle pulse propagation through a
nonlinear metamaterial with higher-order nonlinear dispersion effects and
derive its general traveling wave solution. New types of exact analytical
periodic wave solutions that are composed of the product of pairs of Jacobi
elliptic functions are derived in Sec III. Results for dipole 
soliton solutions is also presented here in the long-wave limit of periodic wave solutions. We also present the exact bright and W-shaped soliton solutions of the model equation. In Sec IV, we introduce the
similarity transformation method for solving the generalized NLSE with varying coefficients. The application of
developed method for the wave-speed management of dipole solitons is presented in Sec V. We also apply the developed technique to the management of bright and W-shaped solitons in Sec VI. Finally, in Sec. VII, we give some concluding remarks.

\section{Generalized NLSE for nonlinear metamaterials}

The generalized NLSE describing the propagation of a few-cycle pulse in
nonlinear metamaterials has the following form for the optical pulse
envelope $U(z,t)$ \cite{M1,M2,M3,M4,M5,M6,M7,M8,M9}, 
\begin{equation}
iU_{z}+i\alpha_{1}U_{t}-\alpha_{2} U_{tt}+\gamma \left\vert U\right\vert
^{2}U=i\lambda (\left\vert U\right\vert ^{2}U)_{t}+i\epsilon (\left\vert
U\right\vert ^{2})_{t}U+\sigma _{1}(\left\vert U\right\vert
^{2}U)_{tt}+\sigma _{2}\left\vert U\right\vert ^{2}U_{tt}+\sigma
_{3}U^{2}U_{tt}^{\ast },  \label{1}
\end{equation}%
where $U(z,t)$ represents the complex envelope of the electrical field, $z$
and $t$ are the propagation distance and time, respectively, while the
parameters $\alpha_{2} ,$ $\gamma ,$ $\alpha_{1} ,$ $\lambda ,$ and $%
\epsilon $ represent the group velocity dispersion, cubic nonlinearity,
intermodal dispersion, self-steepening, and nonlinear dispersion
coefficients, respectively. Also, $\sigma _{i}$ (with $i=1,2,3$) are
higher-order terms that appear in the context of metamaterials.

This equation has gathered significant attention recently for its importance
from various physical view points \cite{M3,M4,M5,M6,M7,M8}. In particular,
the existence of bright and dark soliton solutions of Eq. (\ref{1}) have
been recently investigated using different methods \cite{M1,M9}.
Super-Gaussian envelope solitons \cite{M3} as well as singular solitons \cite%
{M4} have been also presented for this model. Here, we present the important
results showing the wave-speed management of different types of solitons
obtained within the generalized NLSE (\ref{1}) framework. It will be
demonstrated that these soliton modes can be deceleraed and accelerated
through appropriate modification of the distributed parameters.

To start with, we consider the solution of generalized NLSE (\ref{1}) in the
form \cite{Kruglov1,Kruglov2}, 
\begin{equation}
U(z,t)=u(x)\exp [i(\kappa z-\delta t+\theta )],  \label{2}
\end{equation}%
where $u(x)$ is a real amplitude function depending on the traveling
coordinate $x=t-qz$, and $q=v^{-1}$ is the inverse velocity. Also, $\kappa $
and $\delta $ are the respective real parameters describing the wave number
and frequency shift, while $\theta $ represents the phase of pulse at $z=0$.

On substitution of the waveform solution (\ref{2}) into the model (\ref{1}),
one obtains the system of ordinary differential equations:%
\begin{equation}
(q-2\alpha_{2}\delta -\alpha_{1} )\frac{du}{dx}+\left[ 3\lambda +2\epsilon
-2\delta \left( 3\sigma _{1}+\sigma _{2}-\sigma _{3}\right) \right] u^{2}%
\frac{du}{dx}=0,  \label{3}
\end{equation}%
\begin{equation}
\alpha_{2}\frac{d^{2}u}{dx^{2}}+(3\sigma _{1}+\sigma _{2}+\sigma _{3})u^{2}%
\frac{d^{2}u}{dx^{2}}+6\sigma _{1}u\left( \frac{du}{dx}\right) ^{2}-\left[%
\gamma -\lambda \delta +\delta ^{2}\left( \sigma _{1}+\sigma
_{2}+\sigma_{3}\right) \right] u^{3}+(\kappa -\alpha_{1} \delta
-\alpha_{2}\delta ^{2})u=0.  \label{4}
\end{equation}%
In the general case the system of Eqs. (\ref{3}) and (\ref{4}) is
overdetermined because we have two differential equations for the function $%
U(z,t)$. However, if some constraints for the parameters in Eq. (\ref{3})
are fulfilled the system of Eqs. (\ref{3}) and (\ref{4}) has non-trivial
solutions. From Eq. (\ref{3}), one finds the relations: 
\begin{equation}
q=\alpha_{1} +2\alpha_{2} \delta ,~~~~\delta =\frac{3\lambda +2\epsilon }{%
2\left( 3\sigma _{1}+\sigma _{2}-\sigma _{3}\right) }.  \label{5}
\end{equation}%
The latter relations lead to the following expression for the wave velocity $%
v=1/q$, 
\begin{equation}
v=\frac{3\sigma _{1}+\sigma _{2}-\sigma _{3}}{\alpha_{1} (3\sigma
_{1}+\sigma _{2}-\sigma _{3})+\alpha_{2}(3\lambda +2\epsilon )}.  \label{6}
\end{equation}%
Moreover, the relations in Eq. (\ref{5}) reduce the system of Eqs. (\ref{3})
and (\ref{4}) to the ordinary nonlinear differential equation, 
\begin{equation}
(\alpha_{2}+bu^{2})\frac{d^{2}u}{dx^{2}}+cu\left( \frac{du}{dx}\right)
^{2}-du^{3}+\Gamma u=0,~~~~~~~  \label{7}
\end{equation}%
where the parameters $a,$ $b,$ $c$ and $d$ are given by 
\begin{subequations}
\begin{eqnarray}
b &=&3\sigma _{1}+\sigma _{2}+\sigma _{3},~~~~~~~~c=6\sigma _{1},  \label{8}
\\
d &=&\gamma +\frac{(3\lambda +2\epsilon )\left[ 2\epsilon \left( \sigma
_{1}+\sigma _{2}+\sigma _{3}\right) +\lambda (5\sigma _{3}+\sigma
_{2}-3\sigma _{1})\right] }{4(3\sigma _{1}+\sigma _{2}-\sigma _{3})^{2}}, \\
\Gamma &=&\kappa -\frac{(3\lambda +2\epsilon )\left[ 2\alpha_{1} \left(
3\sigma_{1}+\sigma _{2}-\sigma _{3}\right) +\alpha_{2} (3\lambda +2\epsilon )%
\right] }{4(3\sigma _{1}+\sigma _{2}-\sigma _{3})^{2}}.
\end{eqnarray}%
By analytically solving the amplitude equation (\ref{7}), we can identify
the various nonlinear waves that can propagate in the optical metamaterial
described by the generalized NLSE (\ref{1}). Noting that Eq. (\ref{7}) can
be written as 
\end{subequations}
\begin{equation}
\frac{d^{2}u}{dx^{2}}+f(u)\left( \frac{du}{dx}\right) ^{2}+g(u)=0,  \label{9}
\end{equation}%
where the functions $f(u)$ and $g(u)$ are given by 
\begin{equation}
f(u)=\frac{cu}{bu^{2}+\alpha_{2}},~~~~g(u)=\frac{u(\Gamma -du^{2})}{%
bu^{2}+\alpha_{2}}.  \label{10}
\end{equation}%
We introduce new function $Y(u)$ by equation, 
\begin{equation}
\frac{du}{dx}=Y(u).  \label{11}
\end{equation}%
Thus, Eq. (\ref{9}) for new function $Y(u)$ has the following form, 
\begin{equation}
\frac{dY}{du}+f(u)Y+g(u)Y^{-1}=0.  \label{12}
\end{equation}%
Now let us define the function $F(u)$ by relation, 
\begin{equation}
F(u)=Y^{2}(u),  \label{13}
\end{equation}%
then the equation for new function $F$ is 
\begin{equation}
\frac{dF}{du}+2f(u)F+2g(u)=0.  \label{14}
\end{equation}%
The general solution of this equation has the form, 
\begin{equation}
F(u)=F_{0}e^{-G(u)}-2e^{-G(u)}\int_{u_{0}}^{u}g(u^{\prime })e^{G(u^{\prime
})}du^{\prime },  \label{15}
\end{equation}%
where the function $G(u)$ is defined as 
\begin{equation}
G(u)=2\int_{u_{0}}^{u}f(u^{\prime })du^{\prime }.  \label{16}
\end{equation}%
We note the solution in Eq. (\ref{15}) satisfies to the boundary condition
as $F(u_{0})=F_{0}$. In addition, the function $G(u)$ has the following
explicit form, 
\begin{equation}
G(u)=2c\int_{u_{0}}^{u}\frac{u^{\prime }du^{\prime }}{bu^{\prime
2}+\alpha_{2}}=\frac{c}{b}\ln \frac{|bu^{2}+\alpha_{2}|}{|bu_{0}^{2}+%
\alpha_{2}|}.  \label{17}
\end{equation}%
Also, Eqs. (\ref{11}) and (\ref{13}) yield the differential equation: $%
dx=\pm du/\sqrt{F(u)}$. Thus, we find that the general solution of Eq. (\ref%
{7}) takes the form, 
\begin{equation}
x-x_{0}=\pm \int_{u_{0}}^{u}\frac{du^{\prime }}{\sqrt{F(u^{\prime })}},
\label{18}
\end{equation}%
where the function $F(u)$ is defined by Eq. (\ref{15}).

The general solution of Eq. (\ref{7}) can also be written in the form: $%
x=x_{0}\pm \lbrack R(u)-R(u_{0})]$ with $R(u)=\int F^{-1/2}(u)du$. This
solution defines the variable $x$ as a function of the amplitude $u$. Hence,
we have the boundary condition: $x=x_{0}$ at $u=u_{0}$ or $u(x_{0})=u_{0}$.
This general solution leads to many different particular solutions when we
define some constrains to the coefficients of NLSE given in Eq. (\ref{1}).
In what follows, the different classes of periodic (elliptic) waves are
shown to exist in the presence of all physical processes. Such kind of waves
serves as a model of pulse train propagation in optical systems \cite{CDai}.
Moreover, results for dipole, bright and W-shaped soliton solutions are also
obtained for the governing envelope equation.

\section{Periodic and soliton solutions}

\subsection{Periodic waves}

We have found that Eq. (\ref{7}) possesses an exact solution that is
composed of the product of periodic elliptic functions, 
\begin{equation}
u(x)=\pm R\,\mathit{\mathrm{dn}}(w(x-\eta ),k)\,\mathit{\mathrm{s}}\mathrm{n}%
(w(x-\eta ),k),  \label{19}
\end{equation}%
where $\eta $ is an arbitrary constant while $\mathrm{dn}(\zeta ,k)$ and $%
\mathit{\mathrm{sn}}(\zeta ,k)$ are Jacobi elliptic functions of modulus $k$
with $0<k<1\mathrm{.}$ The real parameters $R$ and $w$ in this solution are
defined by the expressions: 
\begin{equation}
R=\sqrt{-\frac{4\alpha_{2} k^{2}}{b}},\quad w=\sqrt{\frac{d}{b(2k^{2}-1)}}.
\label{20}
\end{equation}%
We assume here the constraint as $\sigma _{1}=-(\sigma _{2}+\sigma _{3})/7$.
Further, we obtain the parameter $\kappa $: 
\begin{equation}
\kappa =-\frac{\alpha_{2}d}{b}+\frac{(3\lambda +2\epsilon )\left[2\alpha_{1}
\left( 3\sigma _{1}+\sigma _{2}-\sigma _{3}\right) +\alpha_{2}(3\lambda
+2\epsilon )\right] }{4(3\sigma _{1}+\sigma _{2}-\sigma _{3})^{2}}.
\label{21}
\end{equation}%
Relations (\ref{20}) indicate that the necessary conditions for this
solution to exist are $bd>0$ for $1/\sqrt{2}<k<1$; and $bd<0$ for $0<k<1/%
\sqrt{2}$ and $\alpha_{2}b<0$.

If we insert the solution (\ref{19}) into Eq. (\ref{2}), we obtain a class
of periodic wave solutions for the NLSE model (\ref{1}) of the form, 
\begin{equation}
U(z,t)=\pm R\,\mathit{\mathrm{dn}}(w\xi ,k)\,\mathit{\mathrm{s}}\mathrm{n}%
(w\xi ,k)\exp [i(\kappa z-\delta t+\theta )],  \label{22}
\end{equation}%
where $\xi =t-qz-\eta $ with $\eta $ is the position of wave at $z=0$.

We obtained another periodic solution for Eq. (\ref{7}) that is expressed by
the product of periodic elliptic functions,%
\begin{equation}
u(x)=\pm S\,\mathit{\mathrm{cn}}(p (x-\eta ),k)\,\mathit{\mathrm{s}}\mathrm{n%
}(p (x-\eta ),k),  \label{23}
\end{equation}%
where $\mathit{\mathrm{cn}}(\zeta ,k)$ is Jacobi elliptic function of
modulus $k$\ with $0<k<1$. The real parameters $S$ and $p$ in this solution
are given by 
\begin{equation}
S=\sqrt{-\frac{4\alpha _{2}}{b}},\quad p=\sqrt{\frac{d}{b(2-k^{2})}}.
\label{24}
\end{equation}%
We also assume here the constraint as $\sigma _{1}=-(\sigma _{2}+\sigma
_{3})/7$. Moreover, we obtain the parameter $\kappa $: 
\begin{equation}
\kappa =-\frac{\alpha _{2}d}{b}+\frac{(3\lambda +2\epsilon )\left[ 2\alpha
_{1}\left( 3\sigma _{1}+\sigma _{2}-\sigma _{3}\right) +\alpha _{2}(3\lambda
+2\epsilon )\right] }{4(3\sigma _{1}+\sigma _{2}-\sigma _{3})^{2}}.
\label{25}
\end{equation}

Substitution of (\ref{23}) into (\ref{2}) yields to the following exact
periodic wave solution of Eq. (\ref{1}), 
\begin{equation}
U(z,t)=\pm S\,\mathit{\mathrm{cn}}(p \xi ,k)\,\mathit{\mathrm{s}}\mathrm{n}%
(p \xi ,k)\exp [i(\kappa z-\delta t+\theta )],  \label{26}
\end{equation}%
where $\xi $ is the same as above. From Eq. (\ref{24}), we see that this
periodic solution exists when the following inequalities are satisfied: $%
\alpha _{2}b<0$ and $bd>0$.

We also obtained the periodic solution for Eq. (\ref{7}) that is expressed
by the product of periodic functions as 
\begin{equation}
u(x)=\pm D\,\mathit{\mathrm{cn}}(h(x-\eta ),k)\,\mathit{\mathrm{d}}\mathrm{n}%
(h(x-\eta ),k),  \label{27}
\end{equation}%
The real parameters $D$ and $h$ in this solution are given by 
\begin{equation}
D=\sqrt{\frac{4\alpha_{2} k^{2}}{b(1-k^{2})^{2}}},\quad h=\sqrt{-\frac{d}{%
b(1+k^{2})}}.  \label{28}
\end{equation}%
We also assume here the constraint as $\sigma _{1}=-(\sigma
_{2}+\sigma_{3})/7$. Moreover, we obtain the parameter $\kappa$: 
\begin{equation}
\kappa =-\frac{\alpha_{2}d}{b}+\frac{(3\lambda +2\epsilon )\left[2\alpha_{1}
\left(3\sigma _{1}+\sigma _{2}-\sigma _{3}\right) +\alpha_{2}(3\lambda
+2\epsilon )\right] }{4(3\sigma _{1}+\sigma _{2}-\sigma _{3})^{2}}.
\label{29}
\end{equation}

Substitution of (\ref{27}) into (\ref{2}) yields to the following exact
periodic wave solution of Eq. (\ref{1}), 
\begin{equation}
U(z,t)=\pm D\,\mathit{\mathrm{cn}}(h\xi ,k)\,\mathit{\mathrm{d}}\mathrm{n}%
(h\xi ,k)\exp [i(\kappa z-\delta t+\theta )],  \label{30}
\end{equation}%
where $\xi $ is the same as above. From Eq. (\ref{28}), we see that this
periodic solution exists when the following inequalities are satisfied: $%
\alpha_{2}b>0$ and $bd<0$.

\subsection{Dipole solitons}

Considering the long-wave limit $k\rightarrow 1$, the periodic solutions (%
\ref{22}) degenerates to a dipole-type soliton solution of the form, 
\begin{equation}
U(z,t)=\pm R_{0}~\mathrm{sech}(w_{0}\xi )~\mathrm{th}(w_{0}\xi )\exp
[i(\kappa z-\delta t+\theta )],  \label{31}
\end{equation}
under the parametric condition as $\sigma_{1}=-(\sigma_{2}+\sigma _{3})/7.$
The soliton amplitude $R_{0}$ and the inverse width $w_{0}$ are given by 
\begin{equation}
R_{0}=\sqrt{-\frac{4\alpha_{2} }{b}},\quad w_{0}=\sqrt{\frac{d}{b}},
\label{32}
\end{equation}%
while the wave number $\kappa $ takes the same relation (\ref{21}). Notice
that when $k\rightarrow 1,$ the periodic wave solution (\ref{26}) also
degenerates to the same soliton-type solution (\ref{31}). It follows from
Eq. (\ref{32}) that this dipole soliton exists when the next two conditions
are satisfied: $b\alpha_{2}<0$ and $bd>0$.

For this class of dipole solitons, the energy $E$ is given by 
\begin{equation}
E=\int_{-\infty}^{+\infty}|U(z,t)|^{2}dt= \frac{8|\alpha_{2}|}{3|b|w_{0}}.
\label{33}
\end{equation}%
Notice that the pulse energy $E$ is the integral of motion [$dE/dz=0$] of
the generalized NLSE (\ref{1}) for any optical pulses satisfying the
boundary condition: $U(z,t)\rightarrow 0$ for $t\rightarrow \pm \infty $. An
important observation here is that the energy of dipole solitons depends on
all system parameters which are included in the coefficients $b$ and $d$.
This enable us to change the soliton energy by appropriate manipulation of
the metamaterial parameters.

\subsection{Bright solitons}

We also find that Eq. (\ref{1}) yields a bright soliton solution of the
form, 
\begin{equation}
U(z,t)=\Lambda _{0}~\mathrm{sech}^{2}(w_{0}\xi )\exp [i(\kappa z-\delta
t+\theta )],  \label{34}
\end{equation}%
under the parametric conditions as $\alpha_{2} =0$ and $\sigma_{1}=-(%
\sigma_{2}+\sigma _{3})/7.$ The amplitude $\Lambda _{0}$ of this soliton
solution is a free real parameter while its inverse width and wave number
parameters $w_{0}$ and $\kappa $ are given as 
\begin{equation}
w_{0}=\sqrt{-\frac{d}{2b}},\quad \kappa =\frac{\alpha_{1} (3\lambda
+2\epsilon )}{2(3\sigma _{1}+\sigma _{2}-\sigma _{3})},  \label{35}
\end{equation}%
with the condition $db<0$. Moreover, the corresponding energy $E$ of this
bright soliton reads 
\begin{equation}
E=\int_{-\infty }^{+\infty }|U(z,t)|^{2}dt=\frac{4\Lambda _{0}^{2}}{3w_{0}}.
\label{36}
\end{equation}

\subsection{W-shaped solitons}

In addition, we obtained a W-shaped soliton solution of Eq. (\ref{1}) as
follows:%
\begin{equation}
U(z,t)=B_{0}\left( 1-\frac{3}{2}\mathrm{sech}^{2}(w_{0}\xi )\right) \exp[%
i(\kappa z-\delta t+\theta )],  \label{37}
\end{equation}%
which takes place for the conditions: $\alpha_{2} =0$ and $%
\sigma_{1}=-(\sigma_{2}+\sigma _{3})/7$. The amplitude $B_{0}$ of this
waveform is a free real parameter while the inverse width parameter $w_{0}$
reads 
\begin{equation}
w_{0}=\sqrt{\frac{d}{2b}},  \label{38}
\end{equation}%
with the condition $db>0$. The parameter $\kappa $ for this W-shaped soliton
solution is%
\begin{equation}
\kappa =dB_{0}^{2}+\frac{\alpha_{1}(3\lambda +2\epsilon )}{2(3\sigma
_{1}+\sigma _{2}-\sigma _{3})}.  \label{39}
\end{equation}

\begin{figure}[h]
\includegraphics[width=1\textwidth]{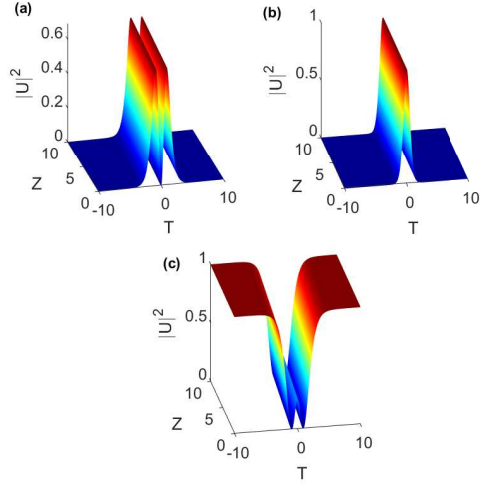}
\caption{Evolution of the soliton solutions with parameters $\protect\sigma%
_{1}=-0.08,$ $\protect\sigma _{2}=0.5,$ and $\protect\sigma _{3}=0.06$ (a)
dipole soliton (\protect\ref{31}) when $\protect\alpha _{1}=1.35,$ $\protect%
\alpha _{2}=-0.25,$ $\protect\gamma =-2,$ $\protect\lambda=0.25,$ $\protect%
\epsilon =0.125,$ $\protect\eta =0$ (b) bright soliton (\protect\ref{34}) $%
\protect\gamma=-1.4075,$ $\protect\alpha _{1}=0.1,$ $\protect\lambda =0.125,$
$\protect\epsilon =-0.25,$ $\protect\eta =0,$ $\Lambda _{0}=1$ (b) W-shaped
soliton (\protect\ref{37}) when $\protect\gamma =-2,$ $\protect\alpha%
_{1}=0.1,$ $\protect\lambda =0.25,$ $\protect\epsilon =0.125,$ $\protect\eta %
=0,$ $B_{0}=1$. }
\label{FIG.1.}
\end{figure}

Figures 1(a),\ 1(b) and 1(c) depict the the evolution of the intensity
profiles of dipole, bright and W-shaped solitons (\ref{31}), (\ref{34}) and (%
\ref{37}) for the parameter values $\sigma _{1}=-0.08,$ $\sigma _{2}=0.5,$
and $\sigma _{3}=0.06$. The other parameters are taken as $\alpha _{1}=1.35,$
$\alpha _{2}=-0.25,$ $\gamma =-2,$ $\lambda =0.25,$ $\epsilon =0.125$ for
solution (\ref{31}), $\gamma =-1.4075,$ $\alpha _{1}=0.1,$ $\lambda =0.125,$ 
$\epsilon =-0.25,$ $\Lambda _{0}=1$ for solution (\ref{34}), and $\gamma
=-2, $ $\alpha _{1}=0.1,$ $\lambda =0.25,$ $\epsilon =0.125,$ $B_{0}=1$ for
solution (\ref{37}). Also, we choose the position $\eta $ of pulses at $z=0$
to be equal to $0.$ \ From these figures, we observe that the soliton
profiles remain unchanged during evolution.

\section{Method of solving generalized NLSE with variable coefficients}

A natural situation would be to consider the inhomogeneities of media in
describing more realistic phenomena. The presence of such inhomogeneities
modifies the propagation dynamics of nonlinear waves since in this case, the
system parameters become functions of the propagation distance. Thus, the
ultrashort light pulse transmission in a realistic nonlinear metamaterial
can be described by the generalized NLSE with varied coefficients:%
\begin{equation}
i\psi _{z}+iD_{1}(z)\psi _{t}-D_{2}(z)\psi _{tt}+R(z)\left\vert \psi
\right\vert ^{2}\psi =i\rho (z)(\left\vert \psi \right\vert ^{2}\psi
)_{t}+if(z)(\left\vert \psi \right\vert ^{2})_{t}\psi +\chi
_{1}(z)(\left\vert \psi \right\vert ^{2}\psi )_{tt}+\chi _{2}(z)\left\vert
\psi \right\vert ^{2}\psi _{tt}+\chi _{3}(z)\psi ^{2}\psi _{tt}^{\ast }.
\label{40}
\end{equation}

\noindent where $D_{1}(z),$ $D_{2}(z),$ $R(z),$ $\rho (z)$ and $f(z)$ are
the variable intermodal dispersion, group velocity dispersion, Kerr
nonlinearity, self-steepening, and nonlinear dispersion coefficients,
respectively. The functions $\chi _{i}(z)$ for $i=1,2,3$ stand for the
varying higher-order nonlinear dispersion coefficients.

Equation (\ref{1}) is an important generalization of the model (\ref{1})
which may be useful for further understanding of the general behavior of the
realistic system. In what follows, we introduce a powerful similarity
transformation method which enables us to study the soliton dynamics in a
real optical metamaterial described by Eq. (\ref{40}). One notes that
studying the control and manipulation of the soliton pulses within the
framework of variable-coefficient NLSE models may help to manage them
experimentally not only in inhomogeneous nonlinear metamaterials but also in
optical fibers and Bose-Einstein condensates where the NLS family of
equations can be also applied to describe the field dynamics.

To obtain the exact analytical solutions for Eq. (\ref{40}), we developed
the method based on the following form of the wave function $\psi (z,t)$:%
\begin{equation}
\psi (z,t)=u(y)\exp [i(\kappa Z(z)-\delta T(z,t)+\theta )],  \label{41}
\end{equation}%
\begin{equation}
y=T(z,t)-qZ(z),~~~~q=\alpha _{1}+2\alpha _{2}\delta ,  \label{42}
\end{equation}%
where $T\left( z,t\right) $ and $Z\left( z\right) $ are two functions to be
determined and the frequency shift $\delta $ is given in Eq. (\ref{5}). It
is important that we present here the function $T\left( z,t\right) $ and
variable $y$ in the special form as 
\begin{equation}
T(z,t)=t+P(z),~~~~y=t-[qZ(z)-P(z)].  \label{43}
\end{equation}%
Moreover, we require that the wave function $\psi (z,t)$ in (\ref{41})
coincides with the wave function $U(z,t)$ in (\ref{2}) for $z=0$. This
initial condition for the wave functions $\psi (z,t)$ and $U(z,t)$ yields
the following initial conditions for the functions $Z(z)$ and $T(z,t)$: $%
Z(0)=0$ and $T(0,t)=t$. Hence, using the expressions in Eq. (\ref{43}) we
have the initial conditions for the functions $Z(z)$ and $P(z)$: $Z(0)=0$
and $P(0)=0$.

Substitution of the expressions (\ref{41}) and (\ref{43}) into Eq. (\ref{40}%
) leads to the following equations: 
\begin{equation}
(P^{\prime }-qZ^{\prime }+D_{1}+2\delta D_{2})\frac{du}{dy}=\left[ 3\rho
+2f-2\delta \left( 3\chi _{1}+\chi _{2}-\chi _{3}\right) \right] u^{2}\frac{%
du}{dy},  \label{44}
\end{equation}%
\begin{equation}
D_{2}\frac{d^{2}u}{dy^{2}}+(3\chi _{1}+\chi _{2}+\chi _{3})u^{2}\frac{d^{2}u%
}{dy^{2}}+6\chi _{1}u\left( \frac{du}{dy}\right) ^{2}-\left[ R-\rho \delta
+\delta ^{2}\left( \chi _{1}+\chi _{2}+\chi _{3}\right) \right]
u^{3}+(\kappa Z^{\prime }-\delta P^{\prime }-\delta D_{1}-\delta
^{2}D_{2})u=0,  \label{45}
\end{equation}%
where $P^{\prime }=dP/dz$ and $Z^{\prime }=dZ/dz$. Eq. (\ref{44}) is
satisfied when the following relations are accepted: 
\begin{equation}
\frac{dP}{dz}-q\frac{dZ}{dz}+D_{1}+2\delta D_{2}=0,  \label{46}
\end{equation}%
\begin{equation}
3\rho +2f-2\delta \left( 3\chi _{1}+\chi _{2}-\chi _{3}\right) =0.
\label{47}
\end{equation}%
Now we perform the important transformation in Eq. (\ref{45}) using the
following change of the variable: $y\mapsto x$. This mapping also yields the
change of function in (\ref{45}) as $u(y)\mapsto u(x)$. One can see that
above mapping does not change the form of function $u$. Moreover, Eq. (\ref%
{45}) is equivalent to Eq. (\ref{4}) when the above mapping of variable is
performed ($y\mapsto x$) and the following equations are satisfied: 
\begin{equation}
\frac{3\chi _{1}+\chi _{2}+\chi _{3}}{D_{2}}=\frac{3\sigma _{1}+\sigma
_{2}+\sigma _{3}}{\alpha _{2}},~~~~\frac{\chi _{1}}{D_{2}}=\frac{\sigma _{1}%
}{\alpha _{2}},  \label{48}
\end{equation}%
\begin{equation}
\frac{R-\rho \delta +\delta ^{2}\left( \chi _{1}+\chi _{2}+\chi _{3}\right) 
}{D_{2}}=\frac{\gamma -\lambda \delta +\delta ^{2}\left( \sigma _{1}+\sigma
_{2}+\sigma _{3}\right) }{\alpha _{2}},  \label{49}
\end{equation}%
\begin{equation}
\frac{1}{D_{2}}\left( \kappa \frac{dZ}{dz}-\delta \frac{dP}{dz}-\delta
D_{1}-\delta ^{2}D_{2}\right) =\frac{\kappa -\alpha _{1}\delta -\alpha
_{2}\delta ^{2}}{\alpha _{2}}.  \label{50}
\end{equation}

By solving Eqs. (\ref{46}-\ref{50}) with the initial conditions $Z(0)=0$ and 
$P(0)=0$ self-consistently, we obtain the following results that define the
mapping variables:%
\begin{eqnarray}
&&\left. P(z)=\frac{\alpha_{1}}{\alpha_{2}} \int_{0}^{z}D_{2}(s)ds-%
\int_{0}^{z}D_{1}(s)ds,\quad Z(z)=\frac{1}{\alpha_{2}}%
\int_{0}^{z}D_{2}(s)ds,\right.  \label{51} \\
&&\left. R(z)=\frac{\gamma}{\alpha_{2}}D_{2}(z),\quad \rho(z)=\frac{\lambda}{%
\alpha_{2}}D_{2}(z),\quad f(z)=\frac{\epsilon}{\alpha_{2}}D_{2}(z), \right.
\label{52} \\
&&\left. \chi _{1}(z)=\frac{\sigma_{1}}{\alpha_{2}}D_{2}(z),\quad \chi
_{2}(z)=\frac{\sigma _{2}}{\alpha_{2} }D_{2}(z),\quad \chi_{3}(z)=\frac{%
\sigma _{3}}{\alpha_{2}}D_{2}(z),\right.  \label{53}
\end{eqnarray}
Thus, all varying parameters in Eq. (\ref{40}) completely defined by
arbitrary function $D_{2}(z)$ via the constraints given in Eqs. (\ref{52})
and (\ref{53}). We also have found that the function $T(z,t)$ and variable $%
y $ have the form, 
\begin{equation}
T(z,t)=t+\frac{\alpha_{1}}{\alpha_{2}} \int_{0}^{z}D_{2}(s)ds-%
\int_{0}^{z}D_{1}(s)ds,  \label{54}
\end{equation}%
\begin{equation}
y=t-\int_{0}^{z}D_{1}(s)ds-2\delta\int_{0}^{z}D_{2}(s)ds.  \label{55}
\end{equation}%
Hence, the dynamics of solitons depends on two arbitrary functions as $%
D_{1}(z)$ and $D_{2}(z)$. Incorporating these results into Eq. (\ref{41}),
we obtain the general self-similar wave solutions to the generalized NLSE
with distributed coefficients (\ref{40}) as%
\begin{equation}
\psi (z,t)=u(t-\zeta(z))\exp[i(\kappa Z(z)-\delta T(z,t)+\theta)],
\label{56}
\end{equation}%
where 
\begin{equation}
\zeta(z)=\int_{0}^{z}D_{1}(s)ds+2\delta\int_{0}^{z}D_{2}(s)ds,  \label{57}
\end{equation}
and parameter $\delta$ is given here in Eq. (\ref{5}). Thus, one can
construct exact self-similar solutions to Eq. (\ref{40}) by using the exact
solutions of Eq. (\ref{1}) via the transformation (\ref{56}). It is worth
noting that the existence of general self-similar solution (\ref{56})
depends on the specific nonlinear and dispersive features of the medium,
which have to satisfy the parametric conditions (\ref{52}) and (\ref{53}).
These conditions present the exact balances among dispersive and nonlinear
processes of different nature.

The equation for velocity of solitons follows from Eq. (\ref{56}) which
yields the relation $dt=(d\zeta /dz)dz$. Thus, the velocity of solitons $%
V(z)= dz/dt$ has the form, 
\begin{equation}
V(z)=\left(\frac{d\zeta(z)}{dz}\right)^{-1}=\frac{1}{D_{1}(z)+2\delta
D_{2}(z)}.  \label{58}
\end{equation}

We emphasize that the solutions described by Eqs. (\ref{56}), (\ref{57}) and
(\ref{51}-(\ref{55}) are satisfied to appropriate boundary conditions. The
substitution into Eqs. (\ref{51}-\ref{53}) the limiting boundary conditions $%
D_{1}(z)=\alpha_{1}$ and $D_{2}(z)=\alpha_{2}$ yield the following
relations: $P(z)=0$, $Z(z)=z$, $R(z)=\gamma$, $\rho(z)=\lambda$, $%
f(z)=\epsilon$, $\chi_{1}(z)=\sigma_{1}$, $\chi_{2}(z)=\sigma_{2}$, $%
\chi_{3}(z)=\sigma_{3}$ which transform NLSE (\ref{40}) into generalized
NLSE (\ref{1}). Moreover, in this case we have $Z(z)=z$, $T(z,t)=t$ and $%
\zeta(z)=(\alpha_{1} +2\alpha_{2}\delta)z=qz$ which leads the wave function
in (\ref{56}) to the form, 
\begin{equation}
\psi (z,t)=u(t-qz)\exp[i(\kappa z-\delta t+\theta)].  \label{59}
\end{equation}%
Hence, in this limiting case we have $\psi (z,t)=U(z,t)$ where the wave
function $U(z,t)$ is defined in (\ref{2}), and the velocity in (\ref{58}) is 
$V(z)=1/q=v$. Note that from this limiting case we have the relation $\psi
(0,t)=U(0,t)$ which is connected with our initial conditions as $Z(0)=0$ and 
$P(0)=0$.

We also note that the developed method of solving the generalized NLSE with
variable coefficients is applied here to Eq. (\ref{40}). However, this
general method can be used for solving an arbitrary NLSE with variable
coefficients. We emphasize that this method for solving the generalized NLSE
with variable coefficients is significantly differ from the mapping given in
the previous papers \cite{Kruglov3,Kruglov4,Dai2,Krug3}.

\section{Wave-speed management of dipole solitons}

Having obtained the general self-similar solution (\ref{56}) of the
physically relevant model (\ref{40}), we now analyze the problem of
wave-speed management of soliton pulses by considering important spatial
modulation of metamaterial parameters. In particular, we will study the
deceleration (i.e., the slowing) and acceleration motions of the dipole
soliton pulse as it propagates through the optical metamaterial.

To start with, we first construct the exact self-similar soliton solutions
of the generalized NLSE with distributed coefficients (\ref{40}).
Substitution of the solution (\ref{31}) into Eq. (\ref{56}) leads to a
family of exact self-similar dipole soliton solution for Eq. (\ref{40}) of
the form,%
\begin{equation}
\psi (z,t)=\pm R_{0}~\mathrm{sech}\left( w_{0}\tau (z,t)\right) \mathrm{th}%
\left( w_{0}\tau (z,t)\right) \exp [i(\kappa Z(z)-\delta T(z,t)+\theta )],
\label{60}
\end{equation}%
where $\tau (z,t)=t-\zeta (z)-\eta $. Here the amplitude $R_{0}$ and inverse
width $w_{0}$ are given in (\ref{32}) and the wave number $\kappa $ in (\ref%
{22}) with the parametric condition as $\sigma _{1}=-(\sigma
_{2}+\sigma_{3})/7$.

Expression (\ref{60}) shows that the amplitude and inverse width of dipole
solitons remain constants as they propagate through the metamaterial while
their velocity $V(z)$, given by Eq. (\ref{58}), is affected by the varying
dispersion coefficients $D_{1}(z)$ and $D_{2}(z).$ This may lead to the
decelerating or accelerating soliton motions by suitable variations of these
parameters. Below, we analyze the deceleration and acceleration processes of
dipole solitons by managing the distributed parameters $D_{1}(z)$ and $%
D_{2}(z)$.

\subsection{Deceleration of a dipole soliton}

To illustrate the wave-speed management of the obtained self-similar dipole
soliton solution (\ref{60}), we consider as example an exponential
distributed control system with a spatial modulation of $D_{1}(z)$ and $%
D_{2}(z)$ in the form,%
\begin{equation}
D_{1}(z)=g_{0}+g_{1}\exp (mz),  \label{61}
\end{equation}%
\begin{equation}
D_{2}(z)=d_{0}+d_{1}\exp (nz),  \label{62}
\end{equation}

\noindent where\ $d_{0},$\ $d_{1},$\ $g_{0},$\ $g_{1}$\ and $m$, $n$ are
real constants. Here, the parameters $m$ and $n$ can be utilized for
adjusting the rate of slowing soliton. The other parameters of the
metamaterial can be obtained exactly through Eqs. (\ref{51}), (\ref{52}) and
(\ref{53}). Consequently, the function $\zeta (z)$ given in Eq. (\ref{57})
can be derived as%
\begin{equation}
\zeta (z)=(g_{0}+2\delta d_{0})z+\frac{g_{1}}{m}\left( e^{mz}-1\right) +%
\frac{2\delta d_{1}}{n}\left( e^{nz}-1\right)  \label{63}
\end{equation}%
It follows from Eq. (\ref{58}) that slowing of dipole soliton can be
achieved when $V(z)\ll 1/q$ or $D_{1}(z)+2\delta D_{2}(z)\gg q$.

\begin{figure}[h]
\includegraphics[width=1\textwidth]{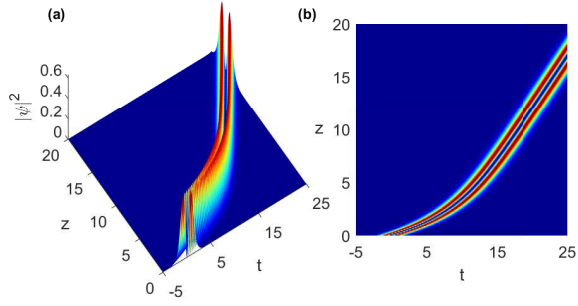}
\caption{Intensity distribution of (a)-(b) the dipole soliton (\protect\ref%
{60}) for $D_{1}(z)$ and $D_{2}(z)$ given by Eqs. (\protect\ref{61}) and (%
\protect\ref{62}). The parameters are $g_{0}=0.5,$ $g_{1}=1.5,$ $d_{0}=0.1,$ 
$d_{1}=0.48,$ $m=-0.5,$ $n=-0.48,$ and $\protect\eta =0$. The other
parameters are the same as in Fig. 1(a). }
\label{FIG.2.}
\end{figure}

Figure 2 depicts the propagation dynamics of the dipole soliton (\ref{60})
in the presence of spatial dependent dispersions $D_{1}(z)$ and $D_{2}(z)$
given by Eqs. (\ref{61}) and (\ref{62}) for the parameter values $g_{0}=0.5,$
$g_{1}=1.5,$ $d_{0}=0.1,$ $d_{1}=0.48,$ $m=-0.5,$ $n=-0.48,$ and $\eta =0$.
The other parameters are chosen as those in Fig. 1(a). One can clearly see
that through this spatial variation of $D_{1}(z)$ and $D_{2}(z)$, the
soliton structure can be slowed as it propagates in the optical
metamaterial. It can be also observed that neither the soliton amplitude nor
width change during the propagation.

\subsection{Acceleration of a dipole soliton}

To accelerate the motion of obtained dipole soliton, we select the
parameters $D_{1}(z)$ and $D_{2}(z)$ to vary periodically along the
propagation direction inside the system as,

\begin{equation}
D_{1}(z)=\frac{gz}{a}\cos (az^{2})-gz^{3}\sin (az^{2}),  \label{64}
\end{equation}%
\begin{equation}
D_{2}(z)=d_{0}+d_{1}\cos (bz),  \label{65}
\end{equation}

\noindent where $g,$ $d_{0}$, $d_{1}$, $a$ and $b$ are constants. The other
metamaterial parameters can be calculated by using Eqs. (\ref{51}), (\ref{52}%
) and (\ref{53}). In this case, the function $\zeta (z)$ can be obtained
using Eq. (\ref{57}) as%
\begin{equation}
\zeta (z)=\frac{1}{2a}gz^{2}\cos (az^{2})+2\delta \left[ d_{0}z+\frac{d_{1}}{%
b}\sin (bz)\right] .  \label{66}
\end{equation}

\begin{figure}[h]
\includegraphics[width=1\textwidth]{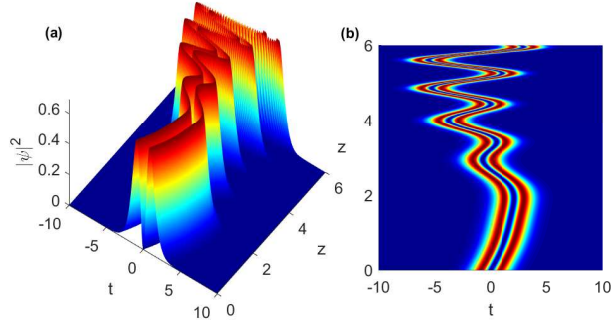}
\caption{Intensity distribution of (a)-(b) the dipole soliton (\protect\ref%
{60}) for $D_{1}(z)$ and $D_{2}(z)$ given by Eqs. (\protect\ref{64}) and (%
\protect\ref{65}). The parameters are $g=-0.3,$ $d_{0}\,=-0.02,$ $d_{1}=1,$ $%
a=1$, $b=1$ and $\protect\eta =0$. The other parameters are the same as in
Fig. 1(a). }
\label{FIG.3.}
\end{figure}

We present the propagation dynamics of the dipole soliton (\ref{60}) with
the distributed parameters $D_{1}(z)$ and $D_{2}(z)$ given by Eqs. (\ref{64}%
) and (\ref{65}) in Fig. 3 with $g=-0.3,$ $d_{0}\,=-0.02,$ $d_{1}=1,$ $a=1$, 
$b=1$, and $\eta =0$. The other parameters are chosen as in Fig. 1(a). We
observe that the dipole soliton displays an accelerating behavior as it
propagates through the optical metamaterial.

\section{Wave-speed management of bright and W-shaped solitons}

Now we turn our attention to discuss the wave-speed management of bright and
W-shaped soliton solutions of the variable-coefficient NLSE (\ref{40}) by
designing appropriate metamaterial parameters. On should note here that
these self-similar wave solutions take place when the constant and variable
parameters of group velocity dispersion vanish (namely, $D_{2}(z)=0$ and $%
\alpha _{2}=0)$. We can find appropriate transformation given in Eqs. (\ref%
{41}) and (\ref{43}) using the found mapping relations in Eqs.(\ref{51}-\ref%
{55}). In this case we make in NLSE (\ref{40}) and other equations the
change $D_{2}(z)\mapsto \alpha _{2}$ and after the limit $\alpha
_{2}\rightarrow 0$. Thus, we have the following mapping in (\ref{40}): $%
R(z)=\gamma $, $\rho (z)=\lambda $, $f(z)=\epsilon $, $\chi _{1}(z)=\sigma
_{1}$, $\chi _{2}(z)=\sigma _{2}$, $\chi _{3}(z)=\sigma _{3}$, and Eqs. (\ref%
{51}) and (\ref{54}) yield 
\begin{equation}
P(z)=\alpha _{1}z-\int_{0}^{z}D_{1}(s)ds,  \label{67}
\end{equation}%
\begin{equation}
Z(z)=z,\quad T(z,t)=t+\alpha _{1}z-\int_{0}^{z}D_{1}(s)ds.  \label{68}
\end{equation}%
We also have the relation $\zeta (z)=\alpha _{1}z-P(z)$ which leads to
equations, 
\begin{equation}
\zeta (z)=\int_{0}^{z}D_{1}(s)ds,\quad V(z)=\left(\frac{d\zeta(z)}{dz}%
\right)^{-1}=\frac{1}{D_{1}(z)}.  \label{69}
\end{equation}%
Moreover, in this case we have the relation $Z(z)=z$, which reduces Eq. (\ref%
{56}) to the form, 
\begin{equation}
\psi (z,t)=u(t-\zeta (z))\exp [i(\kappa z-\delta T(z,t)+\theta )].
\label{70}
\end{equation}

Thus, using the solution (\ref{34}) and the transformation (\ref{70}) we can
write the self-similar bright soliton solution of Eq. (\ref{40}) as 
\begin{equation}
\psi (z,t)=\Lambda _{0}~\mathrm{sech}^{2}\left( w_{0}\tau (z,t)\right) \exp
[i(\kappa z-\delta T(z,t)+\theta )].  \label{71}
\end{equation}%
where $\tau (z,t)=t-\zeta (z)-\eta $. The amplitude $\Lambda _{0}$ is a free
real parameter, the inverse width $w_{0}$ and the wave number $\kappa $ are
given in (\ref{35}) with the parametric conditions as $\alpha _{2}=0$ and $%
\sigma _{1}=-(\sigma _{2}+\sigma _{3})/7$.

The solution (\ref{37}) and transformation (\ref{70}) gives rise to a family
of self-similar W-shaped soliton solution for Eq. (\ref{40}) as%
\begin{equation}
\psi (z,t)=B_{0}\left( 1-\frac{3}{2}\mathrm{sech}^{2}\left( w_{0}\tau
(z,t)\right) \right) \exp [i(\kappa z-\delta T(z,t)+\theta )],  \label{72}
\end{equation}%
where $\tau (z,t)=t-\zeta (z)-\eta $. The amplitude $B_{0}$ is a free real
parameter, the inverse width $w_{0}$ and the wave number $\kappa $ are given
in (\ref{38}) and (\ref{39}) with the parametric conditions as $\alpha
_{2}=0 $ and $\sigma _{1}=-(\sigma _{2}+\sigma _{3})/7$.

\subsection{Deceleration of bright and W-shaped solitons}

Now consider an exponential distributed control system with a spatial
modulation of $D_{1}(z)$ as given by Eq. (\ref{61}). Consequently, the
function $\zeta (z)$ for the bright and W-shaped soliton pulses given in Eq.
(\ref{69}) is determined as 
\begin{equation}
\zeta (z)=g_{0}z+\frac{g_{1}}{m}\left( e^{mz}-1\right) .  \label{73}
\end{equation}%
This shows that the slowing soliton can be achieved for $V(z)=1/D_{1}(z)\ll
1/\alpha _{1}$ or $D_{1}(z)\gg \alpha _{1}.$

\begin{figure}[h]
\includegraphics[width=1\textwidth]{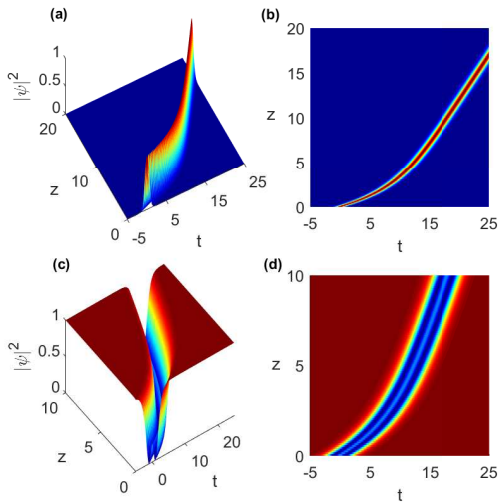}
\caption{Intensity distribution of (a)-(b) the bright soliton (\protect\ref%
{71}) and (c)-(d) the W-shaped soliton (\protect\ref{72}) for $D_{1}(z)$
given by Eq. (\protect\ref{61}). The parameters are $g_{0}=1,$ $%
g_{1}=4,m=-0.5$ and $\protect\eta =0$. The other parameters are the same as
those in Figs. 1(b) and 1(c) respectively. }
\label{FIG.4.}
\end{figure}

In Figs. 4(a)-(b) and 4(c)-(d), we have depicted the propagation dynamics of
the bright and W-shaped solitons (\ref{71}) and (\ref{72}) in the control
system with $D_{1}(z)$ given by Eq. (\ref{61}) for the parameter values $%
g_{0}=1,$ $g_{1}=4,m=-0.5$ and $\eta =0$. The other parameters are chosen as
those in Figs. 1(b) and 1(c) respectively. We clearly see that that through
this spatial variation of $D_{1}(z)$, the soliton pulses can be slowed while
evolving in distance. We also observe that the amplitude and the width of
the solitons keep invariant as distance increases.

\subsection{Acceleration of bright and W-shaped solitons}

We now consider a periodic variation of the intermodal dispersion
coefficient $D_{1}(z)$ in the form (\ref{64}). For this parametric choice,
the function $\zeta (z)$ for the bright and W-shaped solitons given in Eq. (%
\ref{69}) takes the form,%
\begin{equation}
\zeta (z)=\frac{1}{2a}gz^{2}\cos (az^{2}).  \label{74}
\end{equation}

\begin{figure}[h]
\includegraphics[width=1\textwidth]{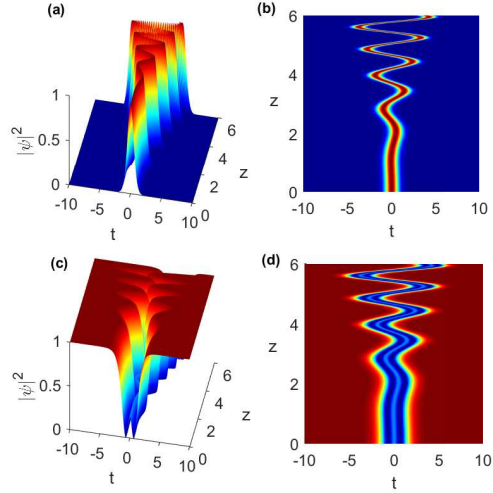}
\caption{Intensity distribution of (a)-(b) the bright soliton (\protect\ref%
{71}) and (c)-(d) the W-shaped soliton (\protect\ref{72}) for $D_{1}(z)$
given by Eq. (\protect\ref{64}). The parameters are $g=-0.3$, $a=1$ and $%
\protect\eta =0$. The other parameters are the same as those in Figs. 1(b)
and 1(c) respectively.}
\label{FIG.5.}
\end{figure}

We present the propagation dynamics of the bright and W-shaped solitons (\ref%
{71}) and (\ref{72}) in the control system with $D_{1}(z)$ given Eq. (\ref%
{64}) in Figs. 5(a)-(b) and 5(c)-(d) with $g=-0.3$, $a=1$ and $\eta =0$. The
other parameters are chosen as those in Figs. 1(b) and 1(c) respectively.
The changes of the soliton trajectories in these figures clearly indicate
their accelerating motion.

\section{Conclusion}

In conclusion, we have studied the wave-speed management of dipole, bright
and W-shaped solitons in an optical metamaterial exhibiting higher-order
dispersive and nonlinear effects of different nature. The propagation of
few-cycle pulse in such system is described by the generalized nonlinear Schr%
\"{o}dinger equation incorporating distributed group velocity dispersion,
cubic nonlinearity, intermodal dispersion, self-steepening, nonlinear
dispersion, and higher-order nonlinear terms that appear in the setting of
metamaterials. We have found that the medium supports the existence of three
new types of periodic wave solutions that are composed of the product of
pairs of Jacobi elliptic functions in the presence of all physical
processes. We have also developed an effective similarity transformation
method to study the soliton dynamics in the presence of the inhomogeneities
of media and analyze the soliton wave-speed management. The results have
showed that an effective control of the soliton wave speed can be achieved
through an appropriate choice of the group velocity dispersion and
intermodal dispersion parameters. In particular, we have demonstrated that
the soliton pulses can be decelerated and accelerated by suitably choosing
the spatial variation of the\ dispersion parameters. We anticipate that the
obtained results will be important in the future investigations of
ultrashort solitons in nonlinear metamaterials.

\end{document}